\documentclass[%
 reprint,
 amsmath,amssymb,
 aps, prl
]{revtex4-1}

\usepackage{graphicx}% Include figure files
\graphicspath{}
\usepackage{float}
\usepackage{upgreek}
\usepackage{dcolumn}% Align table columns on decimal point
\usepackage{bm}% bold math
\usepackage{hyperref}% add hypertext capabilities
\usepackage[mathlines]{lineno}% Enable numbering of text and display math
\usepackage{color}
\usepackage{soul}

 \newcommand{\be}{\begin{equation}}
 \newcommand{\ee}{\end{equation}}
 \newcommand{\bea}{\begin{eqnarray}}
 \newcommand{\eea}{\end{eqnarray}}

\begin{document}

%\preprint{APS/123-QED}

\title{Droplet ejection by electrowetting actuation}% Force line breaks with \\

%\title{Criteria for ejection of satellite droplets during spreading of  droplets on solid substrates by electrowetting}% Force line breaks with \\
%\thanks{Thermal and Fluids Lab, MAE, NTU}%

\author{Quoc Vo}
\author{Tuan Tran}
\email{Corresponding author: xqvo@ntu.edu.sg, ttran@ntu.edu.sg}
\affiliation{School of Mechanical \& Aerospace Engineering, 
	Nanyang Technological University, 
	50 Nanyang Avenue, 639798, Singapore}
\date{\today}% It is always \today, today,             

\begin{abstract}
Fast contact-line motion of 
a droplet spreading on a solid substrate
under the electrowetting effect
generates strong capillary waves 
on the droplet's surface.
%If the electrowetting intensity
%%, controlled 
%%by an applied voltage, 
%is sufficiently high,
The capillary waves
may be strong enough to 
induce ejection of a satellite droplet from the primary one.
%We examine the criteria 
%enabling ejection of satellite droplets.
%In this study, we categorise the ejection
%satellite droplets to two types 
%based on the its observed dynamics, i.e.,
%single-stage ejection 
%and two-stages ejection.
In this study, we show that 
the size of the satellite droplet and 
the ejection time
are not only dependent on the contact-line velocity, 
which directly relates
to the applied voltage enabling the electrowetting effect,
but also affected by 
the ejection dynamics.
We derive 
a theoretical model of the 
criteria for 
droplet ejection 
and experimentally verify the proposed criteria  
for wide ranges of 
viscosity, droplet size 
and the applied voltage.
\end{abstract}

%Use showkeys class option if keyword
\maketitle
\bibliographystyle{apsrev4-1}
Fast motion of contact line 
during spreading of liquid droplets
on solid substrates may result in 
ejection of satellite droplets \cite{Ding2012}.
Such droplet ejection is a 
fascinating physical phenomenon 
involving numerous 
fundamental problems
such as spreading dynamics \cite{Cox1986a}, 
capillary waves \cite{Keller1983a,Billingham1999}, 
pinch-off singularity \cite{Day1998},
coalescence 
\cite{Zhang2015b,Zhang2009,Thoroddsen2000,Shim2017},
or deformation and breakup dynamics 
of double emulsion droplets 
\cite{Chen2013, Chen2015}. 
%or of droplets and a liquid pool \cite{Thoroddsen2000,Shim2017}.}
Understanding the dynamics 
and the criteria
at which ejection happens also
provides important remarks 
for improving industrial processes 
including 
formation of aerosol droplets \cite{Gordillo2019b},
polymer's emulsions, 
industrial sprays \cite{Villermaux2007},
or controlling droplet jumping 
in digital microfluidics \cite{Merdasi2019,Lee2012,Cavalli2016,Hong2015}
and electronics cooling \cite{Foulkes2020}.

The ejection of satellite droplets 
during spreading 
directly ties to
the capillary wave
on the surface 
of the primary droplet \cite{Ding2012}.
In normal droplet wetting phenomena
where the spreading motion 
of a droplet
is driven by 
capillarity at the contact line,
capillary waves on 
the droplet surface
is directly generated from
the fast motion 
of the contact line.
Thus, the ejection of satellite
droplets is only possible 
for high wettability surfaces, 
i.e., those with small static 
contact angles,  
to induce sufficient 
spreading velocity \cite{Ding2012}.
This limits the capacity 
of using normal wetting phenomena
to systematically study 
satellite droplet ejection
during wetting.
Due to this limitation,
%little understanding about 
ejection dynamics 
of satellite droplets during spreading,
as well as the required conditions 
for ejection, have been largely unexplored.
Droplet spreading driven by
electrowetting \cite{Vo2018},
on the other hand, 
does not suffer this limitation.
The velocity of the contact line
of the electrowetting-actuated droplet
is forcefully controlled 
by the electrowetting effect \cite{Vo2021}.
As a result, by using the electrowetting effect
to control droplet spreading,
it is possible to 
trigger droplet ejection
for varying surface 
wettability, liquid
viscosities,
and droplet radius, an advantage that allows us
%to address the current gap of
to systematically investigate ejection phenomena
of satellite droplets.

In this Letter, 
we systematically examine the dynamic of
satellite droplet ejection 
during wetting process
of droplets on solid substrates
in which the spreading velocity
is forcefully controlled 
by the electrowetting effect.
By varying the drop's size, viscosity,
and 
the applied voltage to adjust the electrowetting effect's intensity, 
we study 
the contact line's critical velocity 
 beyond which 
ejection 
of the satellite droplets is possible. 
We also develop a predictive model 
for the condition 
of satellite droplet ejection
and verify it  
experimentally.

 \begin{figure}
\includegraphics[width=0.4\textwidth]{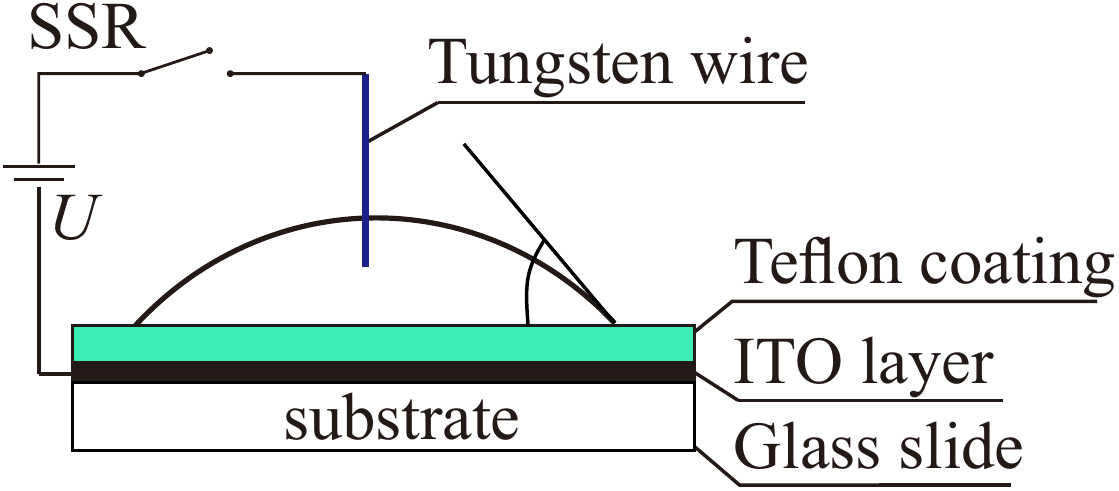}
\centering
\caption{
Schematic of 
the experimental setup using 
electrowetting to 
eject satellite droplets.}
\label{fig:principle}
\vskip -0.5cm
\end{figure}

 \begin{figure*}
\includegraphics[width=0.95\textwidth]{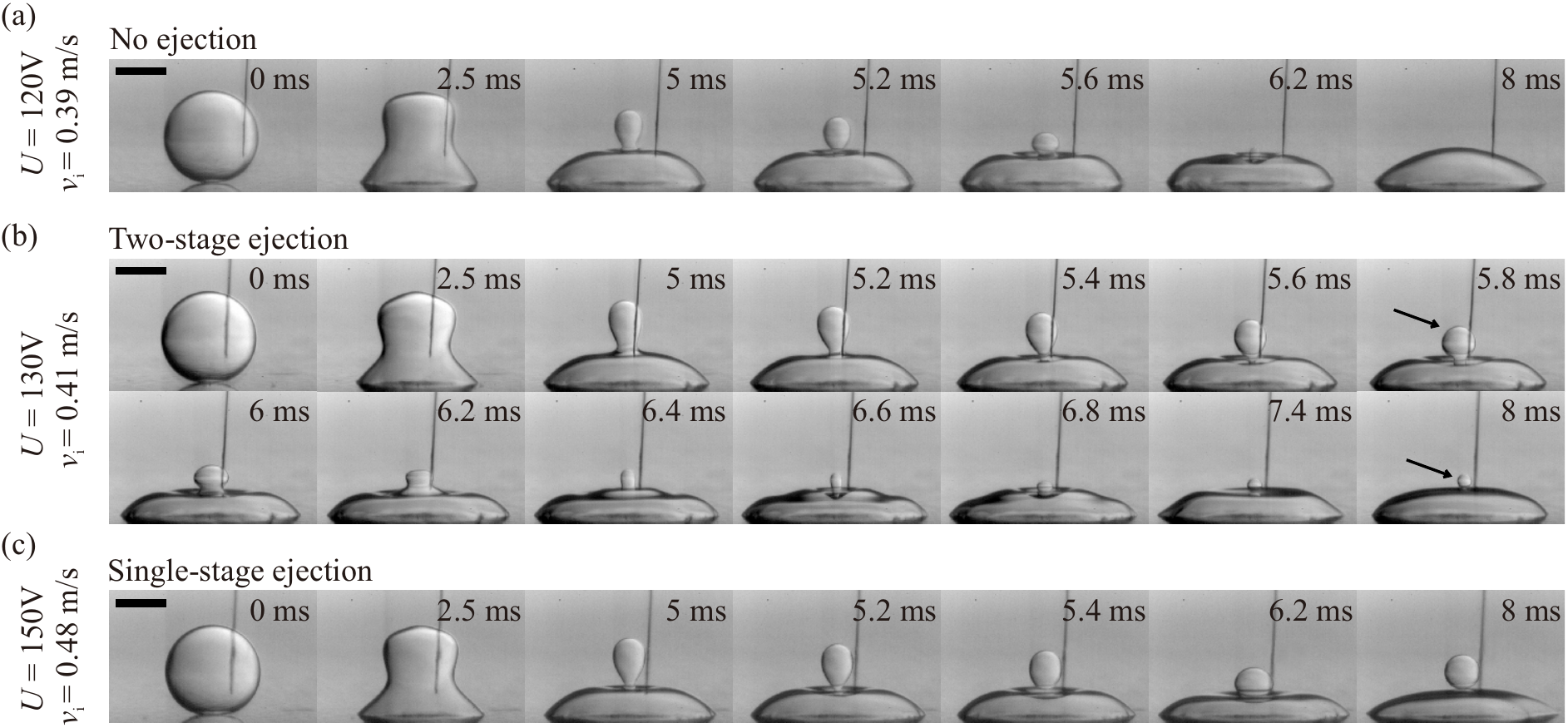}
\centering
\caption{(a) Snapshots showing 
the spreading dynamics of a 
$r_0 =  0.5\,$mm 
droplet when a voltage 
$U = 120\,$V is applied.
No ejection of satellite droplets 
is observed.
(b) Snapshots showing 
the spreading dynamics of a 
$r_0 =  0.5\,$mm droplet
when $U = 140\,$V is applied
resulting in $v_{\rm i} = 0.44\,{\rm m\,s^{-1}}$.
A small satellite droplet is ejected 
when the formed liquid balloon
merges to the droplet.
We categorise this as two-stage ejection.
(c) Snapshots showing 
the spreading dynamics of a 
$r_0 =  0.5\,$mm droplet
when $U = 150\,$V is applied
resulting in $v_{\rm i} = 0.48\,{\rm m\,s^{-1}}$.
A big satellite droplet is ejected directly
from the actuated droplet.
We categorise this as 
single-stage ejection.
In all cases, the scale bars 
represent 0.5\,mm 
and the viscosity of droplets 
is $\eta =  2.2\,$mPa\,s.}
\label{fig:splittingVisual}
\vskip -0.5cm
\end{figure*}

%In our experiment, 
We prepare 
test substrates
to induce electrowetting effect
using indium-tin-oxide (ITO) 
glass slides covered by an amorphous
fluoropolymer (Teflon AF-1601, DuPont) layer. 
%of thickness $d = 2.5 {\rm \upmu m}$.
The ITO layer has thickness of 200\,nm and sheet resistance 20\,$\Omega/{\rm sq}$.
Teflon AF-1601, 
provided by the manufacturer (DuPont)
as a 6\% solution in
Fluorinert® FC-40,
was spincoated 
onto the ITO substrate at the 
speed 1000 round-per-minute 
for 30 seconds using 
a spin coater (Polos Spin150i, APT Automation).
After spincoating,
the coated substrate 
was immediately transferred 
to a hot plate 
for heating process 
to remove the solvent.
The heating process includes
two consecutive steps: 
1) at $115^{\circ}$C 
for $10$ minutes, and 2) 
at $165^{\circ}$C for $30$ minutes \cite{Datta2003}.
The thickness of the Teflon layer is
$d = 2.5\,{\rm \upmu m}$, 
measured using a surface profiler (Talyscan 150, Taylor Hobson).
The Teflon coating has 
surface roughness 
of $\approx 0.6\,$nm 
and negligible variations 
in thickness, 
examined by atomic force microscopy 
(Bioscope Catalyst, Bruker)\cite{Vo2019}.

A droplet is then gently 
deposited on a test substrate 
in every experiment.
To form 
an electrowetting circuit, 
we dip a $18\,\mu$m 
diameter tungsten wire
into the droplet
and connect it to the positive terminal 
of a direct current (DC) power supply (IT6723G, ITECH),
while the ITO layer 
to the negative one
(see Fig.~\ref{fig:principle}). 
We use a solid-state relay (SSR)
to control the application 
of voltage to the electrowetting circuit.
The applied voltage 
is in a form of 
a pulse having amplitude $U$
and pulse's width $T$.
The amplitude $U$
is varied in the range
$0\,{\rm V}\le U \le 190\,$V,
while $T$ is kept fixed 
at $150$\,ms, a
sufficiently 
long period 
to ensure that 
droplets reach new equilibrium
in every actuation.

The working liquids
used to generate droplets 
are solutions of
glycerol, DI water, and 
0.125\,M sodium chloride.
We vary the viscosity $\eta$
of the liquid
from $1\,{\rm mPa\,s}$ 
to $17.6\,{\rm mPa\,s}$
by changing the glycerol mass 
concentrations from 
0\% to 67\% \cite{Vo2018}.
%0\% to 41.5\% \cite{Vo2018}.
The viscosity of the liquid
was measured using a rheometer (Discovery HR-2, TA Instrument).
The radius of the droplets 
$r_0$ is varied 
in the range 
$0.32\,{\rm mm} \le r_0 \le 1.25\,{\rm mm}$.
In every experiment, 
we immerse the droplet 
and the substrate 
in silicone oil 
having viscosity
$\eta_{\rm o} = 1.8\,{\rm mPa s}$ 
and mass density 
$\rho_{\rm o} = 873\,{\rm kg\,m^{-3}}$ (Clearco Products Inc.).
The interfacial tension $\sigma$ between
the working liquid and the oil, 
measured by the pendant drop method, 
varies from 
$37.2\,{\rm mPa\,s}$
to $29.4\,{\rm mPa\,s}$ 
for the tested glycerol solutions \cite{Vo2018}.
The temperature of the oil pool 
is kept at $20 \pm 0.5\,^{\circ}{\rm C}$ 
in all experiments.
We note that all droplets 
in our experiment 
have radii $r_0$
smaller than
the capillary length 
$l_{\rm c} = [\sigma/(\rho - \rho_{\rm o}) g]^{1/2} \approx 5.5\,$mm, 
where $g = 9.781\,{\rm m\,s^{-1}}$ 
is the gravitational acceleration, 
and 
$\rho = 1000\,{\rm kg\,m^{-3}}$ 
is the mass density of the liquid.  
The contact angle of the droplet
on the substrate when no voltage is applied is $\approx 160^{\circ}$.

We use a high speed camera (Photron, SAX2), 
typically running 
at $20000$ frames-per-second, 
to capture the behaviours 
of actuated droplets.
The data presented in 
this Letter are 
obtained by repeating 
experiment with 
the same conditions at least 
three times, 
and the standard deviations of these 
repetitions 
are used for uncertainty estimation.

 \begin{figure*}
\includegraphics[width=0.95\textwidth]{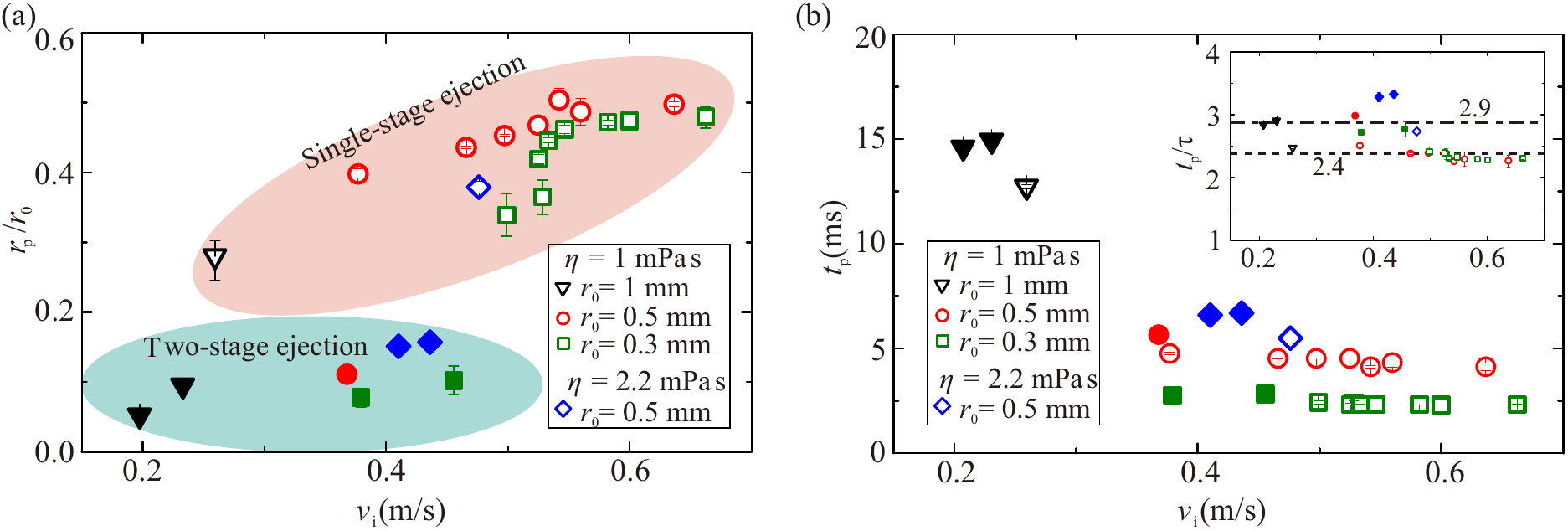}
\centering
\caption{
(a) Plot showing the normalized satellite 
droplets radius $r_{\rm p}/r_0$ 
versus initial 
contact-line velocity $v_{\rm i}$.
The shaded areas 
showing the two groups of $r_{\rm p}/r_0$
corresponding to the two types of droplet ejection
are for guiding the eyes. 
(b) Plot showing the 
pinch-off time $t_{\rm p}$ 
versus $v_{\rm i}$;
$t_{\rm p}$ does not depend on 
$v_{\rm i}$ 
but on $r_0$, $\eta$ 
and types of ejection.
Inset: Plot showing the normalized 
pinch-off time $t_{\rm p}/ \tau$ 
versus $v_{\rm i}$ indicating that 
$t_{\rm p}/ \tau$ 
is well separated by
the types of ejection, i.e., 
$t_{\rm p}/ \tau \approx 2.4$ 
for the single-stage ejection
and $t_{\rm p}/ \tau \approx 2.9$ 
for the two-stage ejection.
In both plots, solid markers are data 
for two-stage ejection,
while open markers are 
for one-stage ejection.}
\label{fig:splittingDinamics}
\vskip -0.5cm
\end{figure*}

In Fig.~\ref{fig:splittingVisual},
we show snapshots presenting
the spreading and 
satellite-droplet ejecting dynamics
of electrowetting-actuated droplets 
with several values of 
the applied voltage $U$.
We observe that, 
when $U$ is applied, 
fast spreading motion of the contact line 
caused by electrowetting effect 
forces the lower part of the droplet
to rapidly expand 
and generates capillary waves 
propagating along 
the droplet-oil interface
toward the apex of the droplet.
When all the waves 
from the circular contact line
hit the apex of the droplet at a same time, 
they create a column of liquid 
on top of the droplet
which subsequently evolves to 
a balloon shape (see Fig.~\ref{fig:splittingVisual}, snapshots at $t \le\,$5\,ms).  
This process is similar to 
the formation of 
pancake shape \citep{Liu2014b} 
or pyramidal shape \citep{Richard2002,Renardy2003} 
of liquid when a
droplet impacts onto a solid substrate.
Increasing the applied voltage $U$ 
leading in 
greater the contact-line spreading velocity 
%presented by a greater value of the
%initial contact-line velocity $v_{\rm i}$
and stronger the capillary waves on 
the droplet's surface 
results in different behaviours 
of the liquid balloon 
and subsequently different
satellite-droplet 
ejection dynamics.
For voltage $U \le 120\,$V, or
equivalently, 
%initial contact-line velocity
$v_{\rm i} \le 0.39\,{\rm ms^{-1}}$,
where $v_{\rm i}$ is the initial value
of the contact-line velocity 
directly controlled 
by the applied voltage \cite{Vo2021},
the liquid balloon is pulled back 
and totally merged 
into the lower part of 
the droplet; 
no ejection of 
satellite droplets occurs
(Fig.~\ref{fig:splittingVisual}a).
%Here, $v_{\rm i}$ is the initial value
%of the contact-line velocity
%which is directly controlled by the applied voltage \cite{Vo2021}.
%Increasing the voltage a bit more,
At $U = 130\,$V 
or $v_{\rm i} = 0.41\,{\rm ms^{-1}}$,
the merging process of 
the balloon 
and the lower part of the droplet
creates 
sufficiently strong capillary waves 
on the balloon surface
and eventually generates 
a small satellite droplet 
on top of the balloon (see Fig.~\ref{fig:splittingVisual}b).
This process is 
similar to the ejection of droplets 
during coalescence of unequal size drops \citep{Zhang2008,Zhang2009}, 
%during coalescence 
of bubbles and droplets \citep{Zhang2015b,Li2014b},
or coalescence cascade 
of a liquid drop \cite{Thoroddsen2000,Shim2017}.
However, due to the high damping of 
our experimental system, 
we did not observe any subsequent coalescence or ejection
of smaller droplets.
We categorise this type of 
droplet ejection as 
two-stage ejection.

At the upper extreme,
i.e., $U \ge 150\,$V
or $v_{\rm i} \ge 0.48\,{\rm ms^{-1}}$, 
the capillary waves on 
the primary droplet
are sufficiently strong
to cause pinch-off 
of the balloon shape 
and eject a much larger 
satellite droplet (see Fig.~\ref{fig:splittingVisual}c).
We categorise this type of ejection 
as single-stage ejection,
a similar process to
the so-called first-stage pinch-off of 
droplets during fast spreading 
reported by Ding et\,al \cite{Ding2012}.

%The two-stage ejection,
%which is similar to 
%the case that droplet ejected 
%during coalescence of
%unequal size droplets \citep{Zhang2009}, 
%is driven by the Laplace 
%pressure difference 
%between the upper and  
%the lower parts of the droplet.
%Larger difference between 
%the two curvatures
%creates stronger downward flows 
%rushing through the neck 
%connecting the upper and lower parts. 
%Subsequently, this creates 
%stronger capillary waves 
%on the droplet's surface.
%As a result, the ejection of 
%the satellite droplets
%depends on the size difference 
%between the two parts of the droplet.
%If we denote $\alpha$ 
%the ratio between the radii
%of the primary droplet 
%and its upper part, 
%the value of $\alpha$
%enabling droplet ejection 
%in our experiment 
%ranges between $2$ and $3$ for 
%Ohnesorge number 
%${\rm Oh} = \eta (\rho \sigma r_0)^{-1/2}$
%varying from $0.0073$ to $0.016$, 
%consistent with the observation 
%for the pinch-off criteria
%of unequal size 
%droplets coalescence
%\cite{Zhang2009}.
%In our experiment, 
%the two-stage ejection
%occurs in a small range 
%of $v_{\rm i}$
%sandwiched between 
%no-ejection and 
%the single-stage ejection.
%For example, with water 
%droplet having $r_0 = 0.5\,$mm, 
%we observe no-ejection for 
%$v_{\rm i} < 0.35\,{\rm ms^{-1}}$,  
%two-stage ejection for
%$0.35\,{\rm ms^{-1}} \le v_{\rm i} < 0.38\,{\rm ms^{-1}}$,
%and single-stage ejection  
%for $v_{\rm i} \ge 0.38\,{\rm ms^{-1}}$.
%As a result, two-stage ejection 
%can be viewed as 
%an intermediate regime
%between no-ejection 
%and single-stage ejection
%when increasing 
%$v_{\rm i}$.

\emph{Size of ejected droplets and ejection time:}
The size of the ejected droplets $r_{\rm p}$ 
and the ejection time $t_{\rm p}$ 
depend on the 
types of droplet ejections,
i.e., single-stage or two-stage.
In Fig.~\ref{fig:splittingDinamics}a,
we show the normalised 
radius of ejected droplets,
$r_{\rm p}/r_0$, versus 
the initial contact-line 
velocity $v_{\rm i}$
for several values of 
droplet radius 
$r_0$ and viscosity $\eta$.
We observe that 
the data are well separated 
depending on the 
type of droplet ejection:
single-stage ejection
possibly results in 
$0.3 \le r_{\rm p}/r_0 \le 0.5$,
whereas 
two-stage ejection 
possibly results in  
$0.05 \le r_{\rm p}/r_0 \le 0.2$.
For both types of ejection, 
$r_{\rm p}/r_0$
increases with $v_{\rm i}$.
However,  
in the single-stage ejection,
$r_{\rm p}/r_0$ increases 
at a higher rate
compared to that 
in the two-stage ejection.
In Fig.~\ref{fig:splittingDinamics}b,
we show the ejection time $t_{\rm p}$
versus 
%the contact 
%line velocity 
$v_{\rm i}$.
We observe that 
$t_{\rm p}$ 
is independent of $v_{\rm i}$ 
for fixed other parameters, i.e., $\eta$, $r_0$
and type of ejection.
There is a measurable 
increase in ejection time
when the viscosity is 
increased from $\eta = 1\,$mPa\,s
to $\eta = 2.2\,$mPa\,s.
We note that it is, however, 
not possible to eject 
satellite droplets 
for viscosity larger 
than $\eta = 2.2\,$mPa\,s.
Therefore, we are not able to examine 
the dependence 
of $t_{\rm p}$ on $\eta$ 
for a larger range of viscosity.
In the inset of 
Fig.~\ref{fig:splittingDinamics}b,
we show the dependence of $t_{\rm p}/\tau$ 
on $v_{\rm i}$, 
where $\tau = (\rho r_0^3/\sigma)^{1/2}$,
revealing that 
for a fixed value of 
viscosity ($\eta = 1\,$mPa\,s),
$t_{\rm p}/\tau \approx 2.42 \pm 0.16$
for one-stage ejection (open markers), 
significantly different 
from that for the two-stage ejection 
($t_{\rm p}/\tau \approx 2.9 \pm 0.14$, solid markers).
The difference in the 
ejection time between two
types of ejection is accounted for
the merging process in
the two-stage ejection (see Fig.~\ref{fig:splittingVisual}).

\emph{Criteria for two-state ejection:}
%which is 
Similar to 
the case that droplet ejected 
during coalescence of
unequal size droplets \citep{Zhang2009}, 
the two-stage ejection
is driven by the Laplace 
pressure difference 
between the upper part, 
i.e., the liquid balloon (Fig.~\ref{fig:splittingVisual}b), and  
the lower part of the droplet.
Larger difference between 
the two curvatures
creates stronger downward flows 
rushing through the neck 
connecting the upper and lower parts. 
Subsequently, this creates 
stronger capillary waves 
on the droplet's surface.
As a result, the ejection of 
the satellite droplets
depends on the size difference 
between the two parts of the droplet.
If we denote $\alpha$ 
the ratio between the radii
of the primary droplet 
and its upper part, 
the value of $\alpha$
enabling droplet ejection 
in our experiment 
ranges between $2$ and $3$ for 
Ohnesorge number 
${\rm Oh} = \eta (\rho \sigma r_0)^{-1/2}$
varying from $0.0073$ to $0.016$, 
consistent with the observation 
for the pinch-off criteria
of unequal size 
droplets coalescence
\cite{Zhang2009}.
In our experiment, 
the two-stage ejection
occurs in a small range 
of $v_{\rm i}$
sandwiched between 
no-ejection and 
the single-stage ejection.
For example, with water 
droplet having $r_0 = 0.5\,$mm, 
we observe no-ejection for 
$v_{\rm i} < 0.35\,{\rm ms^{-1}}$,  
two-stage ejection for
$0.35\,{\rm ms^{-1}} \le v_{\rm i} < 0.38\,{\rm ms^{-1}}$,
and single-stage ejection  
for $v_{\rm i} \ge 0.38\,{\rm ms^{-1}}$.
As a result, two-stage ejection 
can be viewed as 
an intermediate regime
between no-ejection 
and single-stage ejection
when increasing 
$v_{\rm i}$.

\emph{Criteria for single-state ejection:}
%As it was shown in previous studies 
%\citep{Ding2012, Zhang2009},
%the ejection behaviours of 
%droplets by rapid spreading 
%or droplet coalescence
%is governed by 
%the strong capillary waves on 
%the primary droplet's surface
%and resisted by viscous effect.
In the single-stage ejection, 
%of 
%droplet by electrowetting,
the capillary waves
on the droplet's surface
determining the ejection dynamics 
is directly driven by 
the fast motion of 
the contact line with 
the initial velocity $v_{\rm i}$,
and decays over time
by viscous damping 
with the coefficient 
$\xi = \mu (\rho \sigma r_0)^{-1/2}$ \cite{Vo2021}, where $\mu$ is the contact 
line friction coefficient \cite{Vo2018a,Vo2018}.
The necessary condition for the ejection 
to happen is capillary-wave generation at
the droplet-oil interface,
which is previously reported as 
$\xi \le 1$ and ${\rm We} \ge 1$ \cite{Vo2021}.
Here, ${\rm We} = v_{\rm i}^2 \rho r_0/\sigma$ is the 
Weber-like dimensionless 
number 
presenting the ratio of 
inertia caused
by the contact line 
velocity $v_{\rm i}$ 
to the capillarity  
%$v_{\rm i}$ 
%to the inertial-capillary velocity
%$(\sigma/\rho r_0)^{1/2}$
\cite{Vo2021,Keller1983a,Billingham1999}.

We now seek for the sufficient condition 
to eject the satellite from the primary droplet, 
i.e., pinching off at the neck connecting
the satellite droplet and
the lower liquid body (Fig.~\ref{fig:splittingVisual}). 
At the moment pinch-off happens,
the downward velocity $v_{\rm p}$
of the lower liquid body at the neck
must be higher than that %(merging velocity)
of the satellite droplet.
On the one hand,
as fluid motion in the upper neck 
toward the 
lower one is driven 
by the difference in Laplace pressure 
between the satellite droplet 
and the lower body, 
the downward velocity 
of the satellite droplet is 
$(\sigma/\rho r_{\rm p})^{1/2}$. % (Bernoulli)
On the other hand, 
the downward velocity of 
the liquid body below the neck 
is determined by 
the phase velocity $v_{\rm p}$ 
of the capillary waves at the same position.
In our case, the phase velocity $v_{\rm p}$
is estimated as
$v_{\rm p} \sim \lambda f_{\rm p}$,
where $\lambda \sim r_0$ 
is the wavelength,
$f_{\rm p} \sim (k_{\rm d}/2 \pi) (v_{\rm i}/r_0)$ the wave
frequency 
attenuated by the viscous effect,
and $k_{\rm d} = 2 \pi (1 - \xi^2)^{1/2}$
the wave number \cite{Vo2021}.
As a result, 
the sufficient condition 
for the transition 
from no-ejection
to single-stage ejection 
is $v_{\rm i} (1 - \xi^2)^{1/2} = C(\sigma/\rho r_{\rm p})^{1/2}$, 
which can be conveniently 
written in the dimensionless form
%\begin{equation}
%\label{eq:EjectCond}
%{\rm We} = C \left(\frac{r_0}{r_{\rm p}} \right)^{1/2} (1 - \xi^2)^{-1/2}.
%\end{equation}
\begin{equation}
\label{eq:EjectCond}
{\rm We} = C^2 \frac{r_0}{r_{\rm p}} (1 - \xi^2)^{-1}.
\end{equation}
Here, $C$ is a constant of order unity
arisen from the dimensional analysis.

 \begin{figure}
\includegraphics[width=0.45\textwidth]{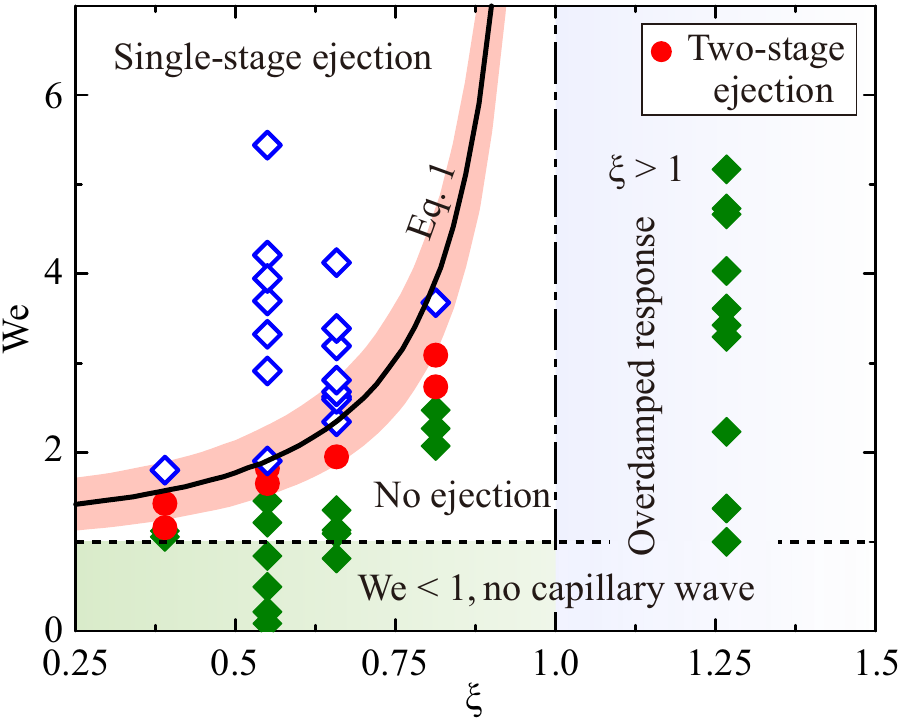}
\centering
\caption{Plot showing 
the phase diagram of 
the ejection behaviours
of water-glycerol droplets 
of various radius $r_0$ and 
viscosity $\eta$ 
rearranged in the dimensionless 
quantities ${\rm We}$
versus $\xi$.
The solid curve presents 
Eq.~\ref{eq:EjectCond} 
using $r_0/r_{\rm p} = 2.5$ 
and $C = 0.73$.
The shaded area around the solid curve
shows the variation 
of the Eq.~\ref{eq:EjectCond}
when the experimental values 
of $r_0/r_{\rm p}$ 
vary from $2$ to $3$.}
\label{fig:splittingPD}
\vskip -0.5cm
\end{figure}

In Fig.~\ref{fig:splittingPD}, 
we show the map of various 
observed behaviours, most notably
the ejection and no-ejection 
behaviours and their transition 
depending on 
two parameters ${\rm We}$ and $\xi$.
The two lines ${\rm We} = 1$ 
and $\xi = 1$
form a coordinate defining 
the quadrant 
in which ejection is possible, 
i.e., the second quadrant 
of the coordinate
where ${\rm We} \ge 1$ 
and $\xi \le 1$.
The region having two-stage ejections 
(red circles)
is clearly sandwiched between
those with no-ejection 
and single-stage ejections.
The transition to single-stage ejection 
is well captured by 
Eq.~\ref{eq:EjectCond} (solid curve).
We note that
since there is no available 
theoretical formula for the values 
of $r_0/r_{\rm p}$,
we use the experimental data 
of $r_0/r_{\rm p}$ 
in Fig.~\ref{fig:splittingDinamics}a 
to test our theory, i.e.,
Eq.~\ref{eq:EjectCond} is 
plotted using the experimental 
value of $r_0/r_{\rm p} = 2.5$ 
%and a prefactor ($C = 0.73$) 
%on the right-hand-side.
and $C = 0.73$.
The shaded area around 
the solid curves in 
Fig.~\ref{fig:splittingPD}
shows the transition 
between no-ejection to ejection regions
caused by the variation 
in the experimental values
of $r_0/r_{\rm p}$ from $2$ to $3$.
The good agreement 
between Eq.~\ref{eq:EjectCond}
and the experimental data
confirms our prediction of 
the condition for 
satellite-droplet ejection by 
fast spreading due to electrowetting. 

\emph{Relation between the applied voltage 
and the ejection criteria:}
With the conditions $\xi \le 1$ 
and ${\rm We} \ge 1$, i.e., 
capillary waves
occur on the surface of 
the electrowetting-actuated droplets, 
one can relate 
${\rm We}$ to $U$ 
by balancing the
electrowetting-induced driving force 
at the contact line 
and the droplet's inertia \cite{Vo2021}:
${\rm We}^{1/2} = A(U - U_{\rm c}) + 1$.
Here, $A$ is 
entirely  
determined by the system's parameters and
$U_{\rm c}$ 
is the voltage  
at which capillary waves 
occur on the droplet's surface.  
We note that it is also possible to 
determine $U_{\rm c}$ theoretically using
capillary waves generation condition  \cite{Vo2021}.
In principle, the relation between ${\rm We}$ and $U$
together with Eq.~\ref{eq:EjectCond} 
provide a complete 
theoretical conditions for satellite 
droplet ejection by electrowetting actuation.

In conclusion, we have ultilized 
the electrowetting effect to induce strong 
capillary waves on droplet-oil interfaces 
and investigated 
the resulting ejection of satellite droplets.
Our results show that 
the radius of ejected droplets
and the ejection time
weakly depend on
the wave magnitude, 
but 
vary substantially 
with the ejection type:
a single-stage ejection 
usually generates larger satellite droplets 
in shorter time 
compared to a two-stage counterpart.
We also experimentally determine 
%Subsequently, we have proposed a model 
%capturing 
the conditions enabling droplet ejection
and proposed a model 
capturing such conditions
%based on 
%two important parameters, 
%i.e., contact-line velocity $v_{\rm i}$ 
%and the viscosity $\eta$
%which are, respectively, 
using 
%nondimensionalized 
%contact-line 
the Weber-like number 
${\rm We}$ 
%of the contact line 
and 
the damping coefficient $\xi$.
An excellent agreement between the theoretical model and the experimental 
data not only 
offers a better understanding in
fundamental problems such as
pinch-off of elongated droplets \citep{Day1998,Burton2007}
or rupture of liquid sheets \citep{Kitavtsev2018},
but also is useful for practical applications 
such as digital microfluidics 
and ink-jet printing.

This study is supported by Nanyang Technological
University and the Agency for Science, 
Technology and Research (A*STAR) under its 
Pharos Funding Scheme (Grant No. 1523700102).
%\vskip -0.5cm

\section{Data Availability}
The data that supports the findings of this study are available within the article.

%\bibliography{/Users/Administrator/Dropbox/MendeleyRefs/JumpingDroplets}
\end{document}